\documentclass[10pt, technote]{IEEEtran}

\usepackage{color}
\usepackage[pdftex]{graphicx}
\usepackage[cmex10]{amsmath}
\usepackage{amsfonts}
\usepackage{cite}
\usepackage{url}
\usepackage{tikz}
\usepackage[normalem]{ulem}
\usepackage{circuitikz}
\usetikzlibrary{calc}
\usepackage{adjustbox}
\usepackage{pdfpages}

\newcommand{\J}{\mathrm{j}}                 

\providecommand{\D}{\,\mathrm{d}}           
\providecommand{\V}[1]{\boldsymbol{#1}}     
\providecommand{\AV}[1]{{\left\langle #1 \right\rangle}_{t_0}} 
\newcommand{\ABS}[1]{\left| #1 \right|}     
\newcommand{\OP}[1]{{\mathcal{#1}}}         

\providecommand{\targ}{\left( t \right)} 
 
\providecommand{\omarg}{\left(\omega\right)}

\providecommand{\ZVAC}{Z_0} 

\providecommand{\Rin}{R_\mathrm{in}}
\providecommand{\Xin}{X_\mathrm{in}}
\providecommand{\Zin}{Z_\mathrm{in}}
\providecommand{\uin}{u_\mathrm{in}}
\providecommand{\uinp}{u_\mathrm{in}^+}
\providecommand{\iin}{i_\mathrm{in}}
\providecommand{\Js}{\V{J}_\mathrm{s}}
\providecommand{\Ji}{\V{J}_\mathrm{i}}

\providecommand{\Plost}{P_\mathrm{lost}}

\providecommand{\Wrec}{\OP{W}_\mathrm{rec}}
\providecommand{\Wlost}{\OP{W}_\mathrm{lost}}
\providecommand{\Wsto}{\OP{W}_\mathrm{sto}}

\providecommand{\Qsto}{Q_\mathrm{sto}}  
\providecommand{\Qrec}{Q_\mathrm{rec}}  
\providecommand{\Qfbw}{Q_\mathrm{FBW}}  

\graphicspath{{figures/}}
\DeclareGraphicsExtensions{.pdf,.jpeg,.png}

\newcommand{\BE}{\begin{equation}}   
\newcommand{\EE}{\end{equation}}     
\newcommand{\BF}{\begin{figure}}     
\newcommand{\EF}{\end{figure}}  	 

\newcommand{\ie}{\textit{i.e.}{}}
\newcommand{\eg}{\textit{e.g.}{}}

\begin{document}
\title{Recoverable Energy of Dissipative Electromagnetic Systems}
\author{Kurt~Schab,~\IEEEmembership{Member,~IEEE}, Lukas~Jelinek, and Miloslav~Capek,~\IEEEmembership{Member,~IEEE}
\thanks{Manuscript received XXX, 2017; revised XXX, 2017.}
\thanks{This work was supported by the Czech Science Foundation under project \mbox{No.~15-10280Y}. Kurt Schab was supported by a grant from the Intelligence Community Postdoctoral Research Fellowship Program.  All statements of fact, opinion, or analysis expressed are those of the author and do not reflect the official positions or views of the Intelligence Community or any other U.S. Government agency.  Nothing in the contents should be construed as asserting or implying U.S. Government authentication of information or Intelligence Community endorsement of the author’s views.}
\thanks{K.~Schab is with the Department of Electrical and Computer
Engineering, Antennas and Electromagnetics Laboratory, North Carolina State University, Raleigh, NC, USA (e-mail: krschab@ncsu.edu).}
\thanks{L.~Jelinek and M.~Capek are with the Department of Electromagnetic Field, Faculty of Electrical Engineering, Czech Technical University in Prague, Technicka~2, 16627, Prague, Czech Republic
(e-mail: \mbox{lukas.jelinek@fel.cvut.cz}, \mbox{miloslav.capek@fel.cvut.cz}).}
}

\markboth{Journal of \LaTeX\ Class Files,~Vol.~6, No.~1, January~2007}%
{XXX}
\maketitle
\begin{abstract}
Ambiguities in the definition of stored energy within distributed or radiating electromagnetic systems motivate the discussion of the well-defined concept of recoverable energy. This concept is commonly overlooked by the community and the purpose of this communication is to recall its existence and to discuss its relationship to fractional bandwidth. Using a rational function approximation of a system's input impedance, the recoverable energy of lumped and radiating systems is calculated in closed form and is related to stored energy and fractional bandwidth. Lumped circuits are also used to demonstrate the relationship between recoverable energy and the energy stored within equivalent circuits produced by the minimum phase-shift Darlington's synthesis procedure.
\end{abstract}

\begin{IEEEkeywords}
Electromagnetic theory, antenna theory, Q~factor, energy storage.
\end{IEEEkeywords}

\section{Introduction}
\label{Intro}

\IEEEPARstart{E}{nergy} stored in the electromagnetic field of circuits \cite{1948_Montgomery_Principles_of_Microwave_Circuits} and antennas \cite{Chu_PhysicalLimitationsOfOmniDirectAntennas,VolakisChenFujimoto_SmallAntennas} has captured the attention of applied physicists for almost a century. This interest is driven largely by the approximate inverse proportionality of the cycle mean stored energy $\langle\Wsto\rangle$ in a time-harmonic steady state and fractional bandwidth of single resonance systems \cite{YaghjianBest_ImpedanceBandwidthAndQOfAntennas,VolakisChenFujimoto_SmallAntennas,IEEEStd_antennas}. This appealing property is, however, hindered by the diversity of available concepts of stored energy $\Wsto\targ$, see \cite{GustafssonJonsson_AntennaQandStoredEnergiesFieldsCurrentsInputImpedance} and references therein, which can lead to extreme differences in values of stored energy (even in cycle mean sense). Notoriously ill-behaving examples include all-pass networks (\eg{}, lossless transmission lines) \cite{1958_Cauer_Synthesis_of_Linear_Communication_Networks}, radiating systems \cite{VolakisChenFujimoto_SmallAntennas} and systems containing dispersive media 
\cite{1976_Barash_Ginzburg_UFN,Loudon_ThePropagationOfElectromagneticEnergyThroughAnAbsorbingDielectric,Ruppin_ElectromagneticEnergy}. In all of these cases, cycle mean stored energy $\langle\Wsto\rangle$ can range from zero to infinity\footnote{One of the widely used methods for calculating the stored energy of radiating systems, \cite{Vandenbosch_ReactiveEnergiesImpedanceAndQFactorOfRadiatingStructures}, is also known to produce negative values of cycle mean stored energy for electrically large  structures \cite{GustafssonCismasuJonsson_PhysicalBoundsAndOptimalCurrentsOnAntennas}.} depending on the convention adopted for defining which part of the total energy in the system is stored. The basis of the aforementioned problems lies in the fact that not all of the internal energy is observable in a lossy system, meaning that its amount cannot be inferred by information available at the input port. In light of this, the stored energy of an electromagnetic system need not be related to port quantities such as fractional bandwidth, though in many practical situations approximate relations are available and are quite accurate \cite{YaghjianBest_ImpedanceBandwidthAndQOfAntennas}.

While the precise definition of stored energy in an electromagnetic system is ambiguous, the maximum energy which can be extracted or recovered from such a system is unique~\cite{1969_Smith_JPRSNSW}. Evaluation of recoverable energy has been derived in a general form \cite{Polevoi_MaximumExtractableEnergy,Direen_FundLimitsOnTerminalBehavourOfAntennas} applicable to any linear electromagnetic system. Recently, an alternative procedure starting from a definition of stored energy \cite{Vandenbosch_RadiatorsInTimeDom1,Vandenbosch_RadiatorsInTimeDom2} has been proposed for the particular case of electric currents in free space \cite{ZhengVandenbosch_RecoverableEnergyICEAA}. Though the procedure described in \cite{Polevoi_MaximumExtractableEnergy} and \cite{Direen_FundLimitsOnTerminalBehavourOfAntennas} is general, concrete examples of its application to antenna systems are missing from the literature.

The aim of this communication is to summarize the concept of recoverable energy, to show its definition, practical means of evaluation, and its relation to both stored energy and fractional bandwidth of a singly-resonant electromagnetic system. The paper is organized as follows. Section~\ref{RecDef} recapitulates a general definition of recoverable energy, while Section~\ref{RecEv} presents practical means of its evaluation. Section~\ref{RecCirc} discusses the relationship between recoverable energy and circuit synthesis.  Section~\ref{Results} shows the numerical results for the recoverable energy of circuits and radiating systems and examines its relationship to their fractional bandwidth. The paper concludes in Section~\ref{Concl}.

\section{Definition of Recoverable Energy}
\label{RecDef}
Although the concept of recoverable energy dates back to the 1970s \cite{1969_Smith_JPRSNSW,1970_Day_QJMAM}, its first explicit appearance within the domain of electromagnetism can be traced to \cite{Polevoi_MaximumExtractableEnergy}, where the recoverable energy $\Wrec$ is defined as
\BE
\label{RecDefEq01}
\Wrec\left( t_0 \right) = \displaystyle\max\limits_{\Js\left(t>t_0\right)} \int\limits_{t_0}^\infty \int\limits_V \V{E} \cdot \Js \D{V}\D{t},
\EE
in which the source current density $\Js$ was used\footnote{The source current density for times \mbox{$t<t_0$} need not be of the same spatial extent as the source current density for times \mbox{$t>t_0$} and varying it allows different values of energy $\Wrec$ to be achieved. In this paper we assume that the extent of the source current density coincides with the extent of a feeding port.} to feed the system for times $t<t_0$. For numerical evaluation of \eqref{RecDefEq01} it is recommended \cite{Polevoi_MaximumExtractableEnergy} to employ Poynting's theorem \cite{Jackson_ClassicalElectrodynamics} which, assuming \mbox{${\V{E}}\left( t = \pm\infty \right) = \V{H}\left( t = \pm\infty \right) = 0$}, claims that the current density $\Js\left( \V{r}, t>t_0 \right)$, which satisfies \eqref{RecDefEq01}, also minimizes the lost energy $\Wlost$, \ie, that the current density $\Js(t>t_0)$ which solves \eqref{RecDefEq01} is also a solution to
\BE
\label{RecDefEq02}
\min_{\Js(t>t_0)}\Wlost\left( t_0, \Js\left( t>t_0 \right) \right).
\EE
The power of \eqref{RecDefEq02} lies in the fact that losses accrued from \mbox{$t = -\infty$} to \mbox{$t=\infty$} are always well defined via \cite{LandauLifshitzPitaevskii1984}
\BE
\label{RecDefEq03}
\begin{split}
\Wlost &\left( t_0, \Js \right) = \int\limits_{-\infty}^\infty  \oint\limits_S \left( \V{E} \times \V{H} \right) \cdot \D{\V{S}}\D{t} \\ &+ \int\limits_{-\infty}^\infty  \int\limits_V \left( \V{E} \cdot \frac{\partial \V{P}_\mathrm{i}}{\partial t} + \V{H} \cdot \frac{\partial \V{M}_\mathrm{i}}{\partial t} + \V{E} \cdot \Ji \right) \D{V}\D{t},
\end{split}
\EE
where subscript $\mathrm{i}$ denotes induced polarizations and, as in \eqref{RecDefEq01} and \eqref{RecDefEq02}, $t_0$ is the time at which energy recovery begins. The only restriction of \eqref{RecDefEq03} is the assumption of linear materials.

In cases when the recoverable energy density within the electromagnetic field \cite{Polevoi_MaximumExtractableEnergy,2007_Glasgow_PRE} is not of primary interest and the system at hand has a well-defined input port of characteristic impedance~$Z_0$, the procedure for obtaining the recoverable energy $\Wrec\left( t_0 \right)$ can further be simplified \cite{Direen_FundLimitsOnTerminalBehavourOfAntennas}. In such a case lost energy can be evaluated as

\BE
\label{RecDefEq04}
\begin{split}
\Wlost\left( t_0, \uinp \left( t>t_0 \right) \right) = \int\limits_{-\infty }^\infty  \uin\targ \iin \targ \D{t} \\ = \frac{1}{{2\pi\ZVAC}} \int\limits_{-\infty }^\infty \ABS{U_\mathrm{in}^+\omarg}^2 \left( 1 - \ABS{\Gamma\omarg}^2 \right) \D{\omega},
\end{split}
\EE
where \mbox{$\uin\targ, \iin\targ$} are the voltage and current at the input port, $\uinp\targ$ is the incident voltage wave \cite{Collin_FieldTheoryOfGuidedWaves}, \mbox{$U_\mathrm{in}^{+}\omarg = \OP{F}\left\{ \uinp\targ \right\}$} is its Fourier transform and $\Gamma\omarg$ is the reflection coefficient \cite{Collin_FieldTheoryOfGuidedWaves}. In this port-oriented approach, the only input variable needed for the evaluation of the energy $\Wrec\left( t_0 \right)$ is the input impedance $\Zin$, as seen from the port.

\section{Evaluation of Recoverable Energy}
\label{RecEv}
In evaluating the recoverable energy $\Wrec$ for circuits and antennas, (\ref{RecDefEq04}) is the most useful. Straightforward manipulations~\cite{Polevoi_MaximumExtractableEnergy} show that \mbox{$\min \left\{ \Wlost\left( t_0,\uinp \right),\uinp\left( t>t_0 \right) \right\}$} is realized by an incident voltage \mbox{$\uinp\targ$} satisfying
\BE
\label{RecDefEq05}
h\targ \ast \uinp\targ  = 0
\EE
for times $t>t_0$, where $\ast$ denotes convolution and
\BE
\label{RecDefEq06}
h\targ = \OP{F}^{-1}\left\{ \displaystyle\frac{1}{\ZVAC} \left( 1-\ABS{\Gamma \omarg}^2 \right) \right\}.
\EE
The integral equation \eqref{RecDefEq05} can be solved in the spectral domain via the Wiener-Hopf method \cite{Polevoi_MaximumExtractableEnergy}, or directly in the time domain by method of moments \cite{Harrington_FieldComputationByMoM}, resulting in the optimal incident voltage \mbox{$\uinp \left( t>t_0 \right)$} for a given incident voltage \mbox{$\uinp \left( t<t_0 \right)$}. Once the optimal time course of $\uinp\targ$ is known, the recoverable energy \mbox{$\Wrec\left( t_0 \right)$} is evaluated as

\BE
\label{RecDefEq07}
\Wrec\left( t_0 \right) = -\int\limits_{t_0}^\infty \uin\targ \iin\targ\D{t},
\EE
where
\BE
\label{RecDefEq08}
\begin{split}
\uin\targ &= \uinp\targ +\uinp\targ \ast \gamma\targ,\\
\iin\targ &= \frac{1}{\ZVAC} \left( \uinp\targ - \uinp\targ \ast \gamma\targ \right)
\end{split}
\EE
in which
\BE
\label{RecDefEq08B}
\gamma\targ = \OP{F}^{-1}\left\{ \Gamma\omarg \right\}.
\EE

In the important case of time-harmonic excitation for times~\mbox{$t<t_0$} of an input impedance \mbox{$\Zin = \Rin + \J \Xin$} in the form of a rational function, the solution to (\ref{RecDefEq05}) can be found analytically \cite[\S 4.4.1]{Direen_FundLimitsOnTerminalBehavourOfAntennas}.
In this procedure, factorization of the squared transmittance function \mbox{$|\kappa|^2 = 1-|\Gamma|^2$} is performed to obtain~$\kappa$ which is analytic in the lower complex half-plane.  Once the rational function representing~$\kappa$ is known, the recoverable energy $\Wrec$ follows from a closed-form expression. For lumped circuits, this procedure is exact. Distributed systems (\eg{}, transmission lines or radiators) exhibit irrational input impedance and, in such cases, this approach is approximate. 

When the input impedance $\Zin$ presents an open-circuit or short-circuit at DC, the transmittance function $\kappa$ naturally has a zero at angular frequency $\omega = 0$. This zero requires special attention during the factorization of $|\kappa|^2$ in order to properly construct a transmittance $\kappa$ which satisfies energy conservation and the appropriate analyticity conditions.

\section{Recoverable Energy and Circuit Synthesis}
\label{RecCirc}

An interesting link exists between recoverable energy $\Wrec$ and a minimum phase-shift\footnote{Not to be confused with classical Darlington's synthesis \cite{Darlington_SurveyOfRealizationTechniques}.} Darlington's synthesis \cite{1969_Smith_JPRSNSW}. If a prescribed input impedance $\Zin$ is represented by a \mbox{positive-real} rational function and an equivalent network is constructed via the minimum phase-shift Darlington's synthesis \cite{1961_Hazony_IRETCT}, it has been shown that the energy stored in its reactive elements is the minimum possible of all equivalent networks \cite{1969_Smith_JPRSNSW} and that this energy is also equal to the recoverable energy $\Wrec$ \cite{Direen_FundLimitsOnTerminalBehavourOfAntennas}. A remarkable aspect of this connection is the necessity to use non-reciprocal elements (gyrators) to represent recoverable energy \cite{Smith_AverageEnergyStorage}, \cite{Direen_FundLimitsOnTerminalBehavourOfAntennas}. The presence of non-reciprocal elements distinguishes the minimum phase-shift Darlington's synthesis from Brune's synthesis which is hypothesized \cite{GustafssonJonsson_AntennaQandStoredEnergiesFieldsCurrentsInputImpedance} to produce equivalent circuits whose storage is the minimum representation of stored energy $\Wsto$. 

As an example of the aforementioned distinction, two circuit representations of the input impedance \mbox{$\Zin = 1+1/\left(1+\J \omega\right)$} \cite{1969_Smith_JPRSNSW} are depicted in Fig.~\ref{fig1:OrigCirc}. The left panel shows a minimum phase-shift Darlington's circuit with energy storage equal to $\Wrec$, while the right panel shows Brune's circuit with a presumably minimum energy storage $\Wsto$ from all reciprocal realizations \cite{GustafssonJonsson_AntennaQandStoredEnergiesFieldsCurrentsInputImpedance}. In particular, taking cycle average at frequency $\omega$, Fig.~\ref{fig1:OrigCirc}a leads to $\AV{\Wsto}=\AV{\Wrec}=\left(3 - 2\sqrt 2\right)/\left(4\left(1+\omega^2\right)\right)$, while Fig.~\ref{fig1:OrigCirc}b leads to $\AV{\Wsto}=1/\left(4\left(1+\omega^2\right)\right)$. Here $\AV{\cdot}$ denotes the cycle mean with respect to $t_0$.\\

\begin{center}
\begin{figure}
		\begin{adjustbox}{scale=0.8}
		\begin{circuitikz}[scale=1]
			\begin{scope}[local bounding box=circA]	
				\coordinate (B1) at (0.5,0.5);
				\coordinate (B2) at (0.5,2.5);
				\draw (B1) to[short, *-o] (0.5,0);
				\draw (B1) -- ++(1,0) to [R, l_=$1\,\Omega$] ++(0,2) -- ++(-1,0);
				\draw (B1) -- ++(-1,0) to [C,l_=$1$\,F] ++(0,2) -- ++(1,0);
				\draw (B2) to[short, *-] ++(0,0) to[R, l_=$1\,\Omega$] ++(0,2) to[short,-o] ++(0,0);
				\draw (-0.8,4.5) node {(b)};
			\end{scope}
			\begin{scope}[shift={($(circA.south east)+(-7cm,+0cm)$)}, local bounding box=circB]	
				\draw (0,2) node[gyrator] (G) {};
                \draw (G.base) node[yshift=-2.0cm] {$\longrightarrow$};
				\draw (G.base) node[yshift=-2.3cm] {$\sqrt{2}\,\Omega$};
				\draw (G.A1) -- ++(0, 0.5) to[short, -*] ($0.5*(G.A1)+0.5*(G.B1)+(0,0.5)$) node(Center){} -| (G.B1);
				\draw (Center) to [C, l_=$\displaystyle\frac{1}{2}$\,F] ++(0,2) to[short, *-o] ++(-1.5,0);
				\draw (G.B2) -- ++(1,0) node(B2R){};
				\draw ($(Center)+(0,2)$) -- ($(G.B1)+(1,2.5)$) to [R=$2\,\Omega$] (B2R);
				\draw (G.A2) to[short, -o] ++(-0.5,0);	
				\draw (-2.5,4.5) node {(a)};                
			\end{scope}		
		\end{circuitikz}
		\end{adjustbox}        
	\caption{Minimum non-reciprocal representation (a) and minimum reciprocal representation (b) of input impedance \mbox{$\Zin = 1+1/\left(1+\J \omega\right)$}.}
	\label{fig1:OrigCirc}	
\end{figure}
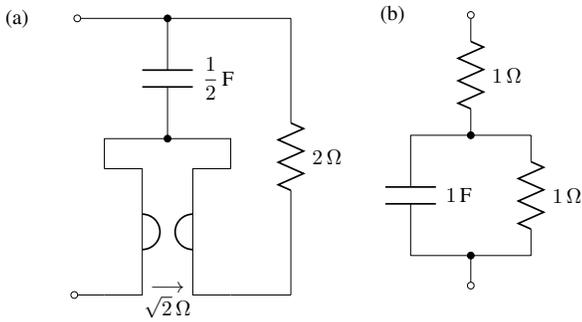
\end{center}

\section{Recoverable Energy of Selected Electromagnetic systems}
\label{Results}

Here the procedure described in Section~\ref{RecEv} is used to evaluate the recoverable energy $\Wrec$ for electromagnetic systems excited to a time-harmonic steady state. For comparison purposes, the stored and recoverable energies are used to compute the quality factors 
\BE
\label{ResultsEq01}
\Qsto = \frac{\omega \AV{\Wsto}}{\Plost}
\EE
and
\BE
\label{ResultsEq02}
\Qrec = \frac{\omega \AV{\Wrec}}{\Plost}.
\EE
These values are in turn compared to the quality factor determined by fractional bandwidth \cite{YaghjianBest_ImpedanceBandwidthAndQOfAntennas}
\BE
\label{ResultsEq03}
\Qfbw = 2\sqrt \frac{\alpha}{1-\alpha} \, \frac{1}{\mathrm{FBW}_\alpha},
\EE
where $\alpha$ denotes a level of $\ABS{\Gamma}^2$ at which the fractional bandwidth FBW is evaluated. In the aforementioned quality factors, the quantity
\BE
\label{ResultsEq04}
{P_{{\mathrm{lost}}}} = \left\langle {\oint\limits_S {\left( {{\boldsymbol{E}} \times {\boldsymbol{H}}} \right) \cdot {\mathrm{d}}{\boldsymbol{S}}} } \right\rangle
\EE
denotes cycle mean lost power \cite{Jackson_ClassicalElectrodynamics} with surface $S$ circumscribing the system at hand, but intersecting the feeding transmission line. Lastly, it is important to stress that in all treated cases, the input impedance is assumed to be tuned at the evaluation frequency $\omega$ by a properly chosen series lumped reactive element \cite{YaghjianBest_ImpedanceBandwidthAndQOfAntennas}, \ie{}, \mbox{$\Xin \left(\omega \right)=0$}.

\subsection{Lumped Circuits}
\label{ResultsA}
The input impedance of a lumped circuit has the form of a positive-real rational function \cite{Wing_ClassicalCircuitTheory} which means that its recoverable energy in a time-harmonic steady state can, in principle, be calculated analytically, see Section~\ref{RecEv}. Particularly simple are the cases of serial and parallel resonance circuits in which it is easy to show \cite{Direen_FundLimitsOnTerminalBehavourOfAntennas} that $\AV{\Wsto}=\AV{\Wrec}$. This is, however, a rare example and in general $\AV{\Wsto}$ is notably greater than~$\AV{\Wrec}$.

As an example, the circuit in Fig.~\ref{fig2} is studied, with values of the quality factors~$\Qsto$, $\Qrec$, and $\Qfbw$ also shown.  For all frequencies, the quality factors $\Qfbw$ and $\Qsto$ are nearly identical, while that given by $\Qrec$ is significantly lower. This difference vanishes for \mbox{$R_1 = 0\,\Omega$} as the circuit becomes a parallel resonance RLC circuit.  The optimal time course of~$\uinp\targ$ for this circuit using a single time $t_0$ and feeding frequency $\omega$ is shown in Fig.~\ref{fig3}.

\subsection{Antennas}
\label{ResultsB}

The input impedance of an antenna cannot generally be expressed as rational function of $\omega$.  Thus, any rational fit used in the analytic recoverable energy calculation described in \cite[\S 4.4.1]{Direen_FundLimitsOnTerminalBehavourOfAntennas} is an approximation. However, a comparison of numerical results obtained from the approximate rational fitting method and a direct Wiener-Hopf solution of (\ref{RecDefEq05}) shows that the effect of this approximation is negligible.  As solving~(\ref{RecDefEq05}) directly via the Wiener-Hopf method is extremely computationally expensive compared to the rational function approximation method, all results presented in this subsection are generated using a rational approximation of input impedance according to the method described in \cite{1999_Gustavsen_TPD}.

Unlike in the case of lumped circuits where stored energy is uniquely defined, calculation of the stored energy in \eqref{ResultsEq01} is problematic for antennas and other distributed electromagnetic systems due to the ambiguities discussed in Section~\ref{Intro}.  For comparison purposes, we adopt the definition of stored energy developed in~\cite{Vandenbosch_ReactiveEnergiesImpedanceAndQFactorOfRadiatingStructures}.  Despite its issues, the method of calculating stored energies from a given current distribution presented in~\cite{Vandenbosch_ReactiveEnergiesImpedanceAndQFactorOfRadiatingStructures} remains one of the most practical methods for approximating the energy stored by perfectly conducting antenna systems.

The quality factors for a cylindrical dipole fed by a delta-gap source at its center, as calculated by stored energy \eqref{ResultsEq01}, recoverable energy \eqref{ResultsEq02}, and fractional bandwidth \eqref{ResultsEq03}, are shown in Fig. \ref{fig4}.  Impedance and current data were generated in FEKO \cite{feko}.  At low frequencies where $ka \ll 1$, all methods agree.  However, for higher frequencies the quality factor calculated from the recoverable energy is significantly less than that obtained from the stored energy or fractional bandwidth.  While not an exact match, the quality factor obtained from the stored energy is very close to that obtained from the fractional bandwidth for all depicted frequencies.  The optimal time course for this antenna using $\uinp\targ$ for a single time $t_0$ and feeding frequency $\omega$ is shown in Fig.~\ref{fig5}.

\BF
\includegraphics[width=9cm]{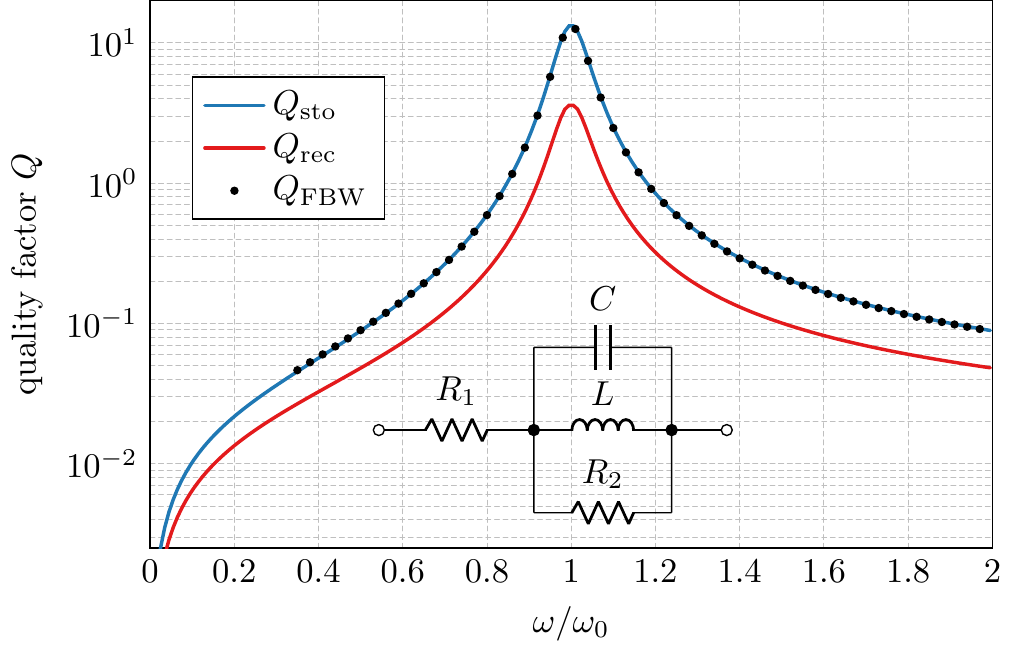}
\caption{Quality factors calculated via energies stored in lumped components~\eqref{ResultsEq01}, recoverable energy~\eqref{ResultsEq02}, and fractional bandwidth~\eqref{ResultsEq03}.  To generate these data, the following component values were used: \mbox{$R_1 = 1\,\Omega$}, \mbox{$R_2 = 2\,\Omega$}, \mbox{$L = 0.1$\,H}, \mbox{$C = 10$\,F}, and \mbox{$\omega_0 = 1$\,s$^{-1}$}.  Fractional bandwidth was calculated using \mbox{$\alpha = 0.0001$}.}
\label{fig2}
\EF

\BF
\includegraphics[width=9cm]{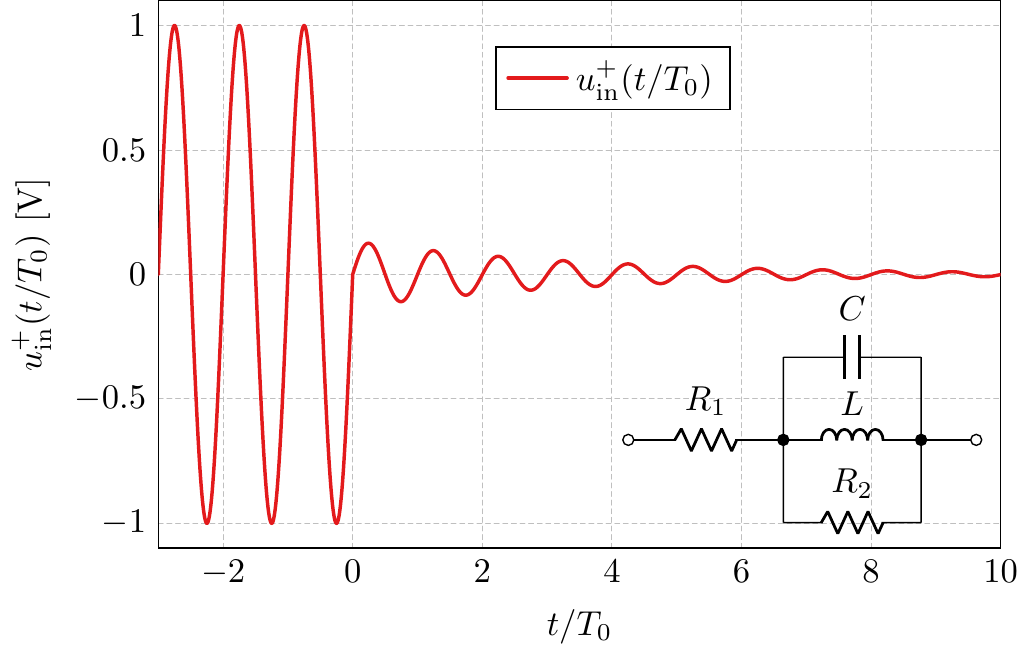}
\caption{Time course of feeding and recovering incident voltage wave $\uinp\targ$ corresponding to the circuit from Fig.~\ref{fig2}. The curve corresponds to time \mbox{$t_0 / T_0 = 0$} and feeding frequency \mbox{$\omega / \omega_0 = 1$.} Normalization of time with respect to \mbox{$T_0 = 2\pi / \omega_0$} is used. Port impedance $Z_0=1\, \Omega$ was used.}
\label{fig3}
\EF

\BF
\includegraphics[width=9cm]{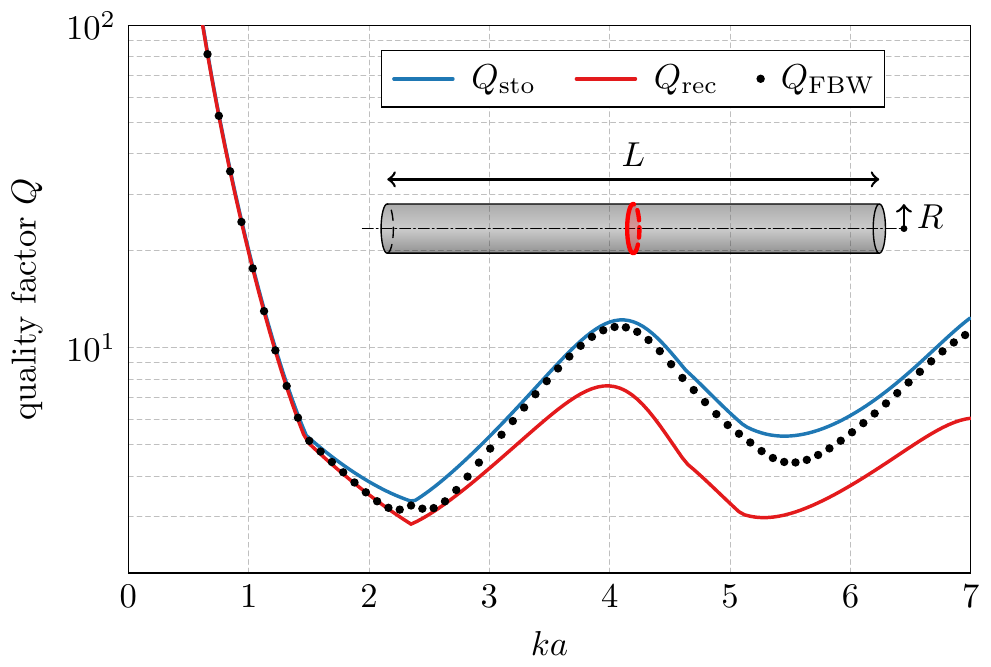}
\caption{Quality factors of a cylindrical dipole with \mbox{$L/R = 200$} calculated via stored energy~\eqref{ResultsEq01}, recoverable energy~\eqref{ResultsEq02}, and fractional bandwidth~\eqref{ResultsEq03}. The driven voltage gap is highlighted by the red ellipse.  Fractional bandwidth was calculated using \mbox{$\alpha = 1/5$}.}
\label{fig4}
\EF

\BF
\includegraphics[width=9cm]{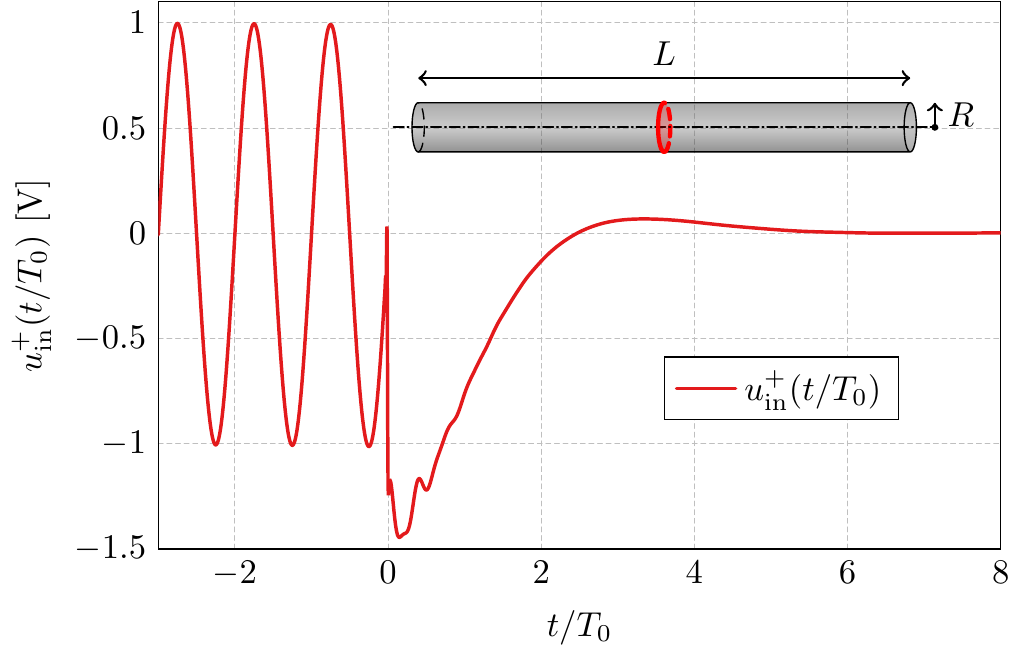}
\caption{Time course of feeding and recovering incident voltage wave $\uinp\targ$ corresponding to the cylindrical dipole from Fig.~\ref{fig4}. The curve corresponds to time \mbox{$t_0 / T_0 = 0$} and feeding frequency $f_0$ coinciding with the first resonance frequency of the dipole. Normalization of time with respect to \mbox{$T_0 = 1/f_0$} is used. Port impedance $Z_0=200\, \Omega$ was used.}
\label{fig5}
\EF

\section{Conclusion}
\label{Concl}
Minimum energy concepts are of paramount importance throughout physics and engineering. One such concept is that of recoverable energy, the value of which determines the minimum possible energy storage in an electromagnetic system with a prescribed input impedance under fixed excitation or, equivalently, the maximum possible energy that can be extracted from the electromagnetic field surrounding it. Due to the relation of energy storage and fractional bandwidth of single resonance systems, the recoverable energy also approximates an upper bound to fractional bandwidth, although the results presented above suggest that stored energy defined by classical means, such as by Brune's synthesis, provides a tighter bound.

The presented communication can be used as a guide through a rather poor publication history of recoverable energy developments and, importantly, as a guide to its practical evaluation. A key observation is that the recoverable energy can be calculated in a straightforward way directly from a system's input impedance.  The text also suggests that although the recoverable energy will not be the first choice for bandwidth estimate of common antennas, its generality can be of great importance in strongly dispersive systems where common ways of evaluating energy storage are problematic.

\section*{Acknowledgment}
The authors would like to thank Mats Gustafsson from Lund University (Sweden) for interesting discussions that stimulated the development of the presented work. We would also like to thank him for sharing the idea behind footnote$^2$, pointing out the notion of minimum energy realizations, and for providing us with rational approximations.
\ifCLASSOPTIONcaptionsoff
  \newpage
\fi

\end{document}